\documentclass[conference]{IEEEtran}
\IEEEoverridecommandlockouts
\usepackage{cite}
\usepackage{amsmath,amssymb,amsfonts}
\usepackage{algorithmic}
\usepackage{graphicx}
\usepackage{textcomp}
\usepackage{xcolor}
\usepackage{hyperref}
\usepackage{array}
\usepackage{booktabs}

\def\BibTeX{{\rm B\kern-.05em{\sc i\kern-.025em b}\kern-.08em
    T\kern-.1667em\lower.7ex\hbox{E}\kern-.125emX}}
\begin{document}

\title{Morse Code-Enabled Speech Recognition for Individuals with Visual and Hearing Impairments}

\author{
    \IEEEauthorblockN{Ritabrata Roy Choudhury}
    \IEEEauthorblockA{
        \textit{School of Computer Engineering}\\
        \textit{Kalinga Institute of Industrial Technology}\\
        Bhubaneswar, Odisha, India \\
        ritabrata2003.rrc@gmail.com \\
    }
}

\maketitle

\begin{abstract}

The proposed model aims to develop a speech recognition technology for hearing, speech, or cognitively disabled people. All the available technology in the field of speech recognition doesn't come with an interface for communication for people with hearing, speech, or cognitive disabilities. The proposed model proposes the speech from the user, is transmitted to the speech recognition layer where it is converted into text and then that text is then transmitted to the morse code conversion layer where the morse code of the corresponding speech is given as the output. The accuracy of the model is completely dependent on speech recognition, as the morse code conversion is a process.  The model is tested with recorded audio files with different parameters. The proposed model's WER and accuracy are both determined to be 10.18\% and 89.82\%, respectively.

\end{abstract}

\begin{IEEEkeywords}
Speech Recognition, Morse Code, Disabled People, Morse Code Dictionary
\end{IEEEkeywords}

\section{Introduction}
\label{sec:introduction}
The process of converting spoken language into text is known as speech recognition. It is a type of technology that enables computers to recognize and respond to spoken commands or questions. Speech recognition is classified into two types: command and control and dictation. Command-and-control systems are intended to recognize specific commands, whereas speech-to-text systems are intended to transcribe spoken language into text. Speech recognition is divided into two different approaches: rule-based and statistical. To interpret speech, rule-based systems use a set of predefined rules and grammar, whereas statistical systems use machine learning algorithms to learn patterns in speech and translate them into text. Statistical systems are more complex than rule-based systems but are also more accurate and flexible. They can adapt to new vocabularies and speaking styles over time and can handle a wide range of accents, dialects, and speaking styles. Speech recognition technology is widely used in various applications, such as voice assistants, voice search, voice-controlled devices, and transcription services. 

There are several drawbacks to existing speech recognition models, including limited accuracy, limited language support, and so on. But one of the major drawbacks of speech recognition is limited accessibility for people with disabilities. People with hearing, speech, or cognitive impairments may find it difficult to use speech recognition models, and they may not be fully accessible to people with disabilities. This is where our model has tried to fill this gap with a technology that will make speech recognition fully accessible to people with disabilities. Our model has incorporated morse code with speech recognition, thereby making the speech(audio) interpretable by disabled people.

Morse code is a method of transmitting text information as a series of on-off tones, lights, or clicks that a skilled listener or observer can directly understand without the use of special equipment. It was invented in the 1830s and 1840s by Samuel Morse and Alfred Vail for use with the telegraph, the first form of electrical communication. The basic Morse code unit is the dot (short signal), which lasts one unit of time, and the dash (longer signal), which lasts three units of time. A unique sequence of dots and dashes represents the letters of the alphabet, numbers, and punctuation marks. For instance, the letter "A" is represented by a single dash, the letter "N" by a single dot followed by a single dash, and the number "1" by a single dot. Morse code was widely used for telegraph communication for many decades, and it remained in use for some time after the invention of the telephone. It is now primarily used by amateur radio operators as a form of signaling and occasionally in emergency situations. Morse code is not a human-readable text.

Morse code is sometimes used as an alternative mode of communication for the challenged of hearing and deaf individuals, or who have difficulty speaking or using their hands. Morse code can also be used as an assistive technology for people with mobility issues, such as those who have difficulty typing on a keyboard with their hands. They can communicate via a morse code interface, such as a simple switch or a sip-puff device activated by blowing or inhaling. In all of these cases, using morse code allows people with disabilities to communicate and interact with the world around them, even when other forms of communication are difficult.

\subsection{Relate Work}

Yang Xu conducted a measuring study using Google+, i-Chat, and Skype \cite{xu2012video}. They looked into these programmers' architectural nuances. The authors revealed some performance information on these applications using passive and active trials, including video generation and adaptation methods, packet loss recovery approaches, and end-to-end delays. According to their research, the location of the server dramatically impacted both user performance and loss recovery in server-based apps. The use of Forward Error Correction (FEC), an error management approach for streaming across unpredictable network connections, was also refuted in favor of batched re-transmissions as a better option for real-time applications.

A measuring investigation on the effectiveness of Skype's Forward Error Correction (FEC) mechanism was carried out by Te-Yuan Huang \cite{huang2010could,huangcould}. They looked at the trade-offs between the quality of the user experience, the redundancy caused by the Forward Error Correction (FEC) mechanism, and the redundancy caused by the Forward Error Correction (FEC) mechanism, as well as the amount of redundancy added by the Forward Error Correction (FEC) process. To maximize the quality of the user experience, they looked for the right amount of redundancy.

A study on Skype's speech rate adaptation in various network scenarios was also carried out by Te-Yuan Huang \cite{huang2009tuning}. The results of this investigation showed that implementing public-domain codecs was not the optimal choice in terms of user satisfaction. To execute their experiment in this study, the researchers took into account various packet loss levels. They then developed a model to manage redundancy under various packet loss scenarios.

A measurement methodology for users' QoE was put forth by Kuan-Ta Chen \cite{chen2009oneclick}. OneClick, the framework they suggested, offered a specific key that users could hit if they weren't happy with the network settings for streaming video. Two applications—instant messaging and shooter games—have OneClick installed. 

Kuan-Ta Chen \cite{chen2009crowdsourceable} offered yet another methodology for quantifying the quality of a user's experience. The proposed technology supported crowd-sourcing since it could validate contributors' inputs. This framework facilitates participation and yields interval-scale scores. They contend that this approach can be used by researchers to gauge user experience quality without degrading the quality of the research findings and to increase user participation diversity at a minimal cost.

Lukasz Budzisz \cite{budzisz2011fair} has conceived and developed a delayed-based congestion control. In homogeneous networks, the suggested system delivers low-standing queues and delays, while in heterogeneous networks, it enables balanced loss- and delay-based flows. They contend that this system outperforms TCP traffic and can accomplish these qualities with a range of loss values. Analyses and experiments show that this system ensures the aforementioned features.Hayes suggested a technique that accepts packet loss unrelated to congestion \cite{hayes2010improved} . They demonstrated experimentally that the suggested approach increases throughput by 150\% with 1\% packet loss and increases capacity sharing by more than 50\%.

Akhshabi suggested doing an experimental analysis of rate adaptation techniques for HTTP streaming \cite{akhshabi2011experimental,akhshabi2012experimental}. Three popular video streaming applications were experimentally assessed over a range of bandwidth ranges. The findings of this study demonstrated that the performance of such streaming applications is not always impacted by TCP congestion management and its reliability requirement. It is yet unknown how rate adaption logic and TCP congestion management interact.

Chen conducted an experimental study to examine how multi-path TCP performed across wireless networks \cite{chen2013measurement}. They calculated the latency brought on by various cellular data carriers. According to the study's findings, Multi-path TCP provides a reliable data transfer under a variety of network traffic scenarios. It should be thought about as a potential expansion of this study to examine the trade-offs between energy costs and performance.

Google is actively developing QUIC (Quick UDP Internet Connections), a new transport protocol for the Internet \cite{assefi2015experimental, assefi2016measuring}. QUIC employs UDP to address packet delay issues in TCP connections with varying packet loss values. QUIC uses multiplexing and FEC to resolve this issue.
Cicco and colleagues carried out an experimental examination on the Google Congestion Control (GCC) in the RTCWeb IETF WG \cite{de2013experimental}. For their experiment, they used a controlled testbed. The experimental investigation findings show that the suggested approach works successfully, however it does not use the bandwidth evenly when it is shared by two GCC flows or a GCC and a TCP flow.

Cicco has also conducted experimental research into Akamai's High Definition (HD) video distributions \cite{de2010experimental , messeri2024artificial}. They gave specifics on the client-server protocol used by Akamai to implement the quality adaptation algorithm. According to their research, the suggested method encodes any video at five distinct bit rates and stores them all on the server. Based on the signal it receives from the silent, the server chooses the bit rate that corresponds to the bandwidth measurement. Depending on the available bandwidth, the bitrate level adapts. The authors of the research also assessed the algorithm's dynamics in three scenarios.

To evaluate the quality of the user experience on television and mobile applications, Winkler \cite{winkler2003video,winkler2003videos} conducted a series of studies. Their suggested experiment takes into account various bitrates, contents, codecs, and network traffic situations.The authors of the research used Double Stimulus Impairment Scale (DSIS) \cite{perez2024guide} and Single Stimulus Continous Quality Evaluation (SSCQE) on the identical collection of materials. They contrasted the two approaches and looked at the experiment outcomes in relation to codec performance.

Oh Hyung Rai \cite{oh2011mesh} suggests mesh-pull-based P2P video streaming with fun-train coding. The recommended method offers streaming that is easy, rapid, and smooth. The proposed system outperforms existing buffer-map-based video streaming systems in packet loss scenarios, according to experimental testing. This study might be extended by looking at jitters as another crucial component and evaluating the behavior of the suggested system in light of jitters values.

Smith \cite{smith2012limit} takes into account the use of Fountain Multiple Description Coding (MDC) in video streaming via diverse peer-to-peer networks. They come to the conclusion that Fountain MDC codes are advantageous in these circumstances, however, there are some limitations in actual P2P streaming systems.

Finally, Vukobratovic \cite{vukobratovic2009scalable,vukobratovic2007scalable} suggested a unique real-time multicast Expanding Window Fountain (EWF) code-based multicast streaming system. Precoding analogous to Raptor has been discussed as a potential advancement in this area.

\subsection{Motivation and Contribution}

In \cite{kumar2012hindi}, Kuldeep Kumar, R.K. Aggarwal, and Ankita Jain develop their own voice recognition system using the Mel frequency cepstral coefficient (MFCC) approach and the hidden Markov model toolkit (HTK). In order to deliver real-time voice-based machine translation, authors examine existing speech recognition technologies. Using Dragon Medical 11.0, Hanna Suominen, Liyuan Zhou, Leif Hanlen, and Gabriela Ferraro \cite{suominen2015benchmarking} propose using voice recognition to prevent errors in information flow in healthcare. In \cite{assefi2016measuring}, Siri, Google Speech Recognizer, and Dragon were used to analyze cloud-based speech recognition systems. Numerous authors contrast their own systems. Using open-source code, Belenko M.V. and Balakshin P.V. \cite{belenko2017comparative} examine systems, enter evaluation coefficients for various parameters, and offer suggestions for recognition systems. 

It is advised to employ HTK and Julius in voice recognition instructional activities. Research tasks can be successfully conducted with Kaldi \cite{belenko2017comparative}. Examine deep neural networks (DNNs) with many hidden layers as well as the methods used to train them in \cite{hinton2012deep}. In \cite{jaitly2012application}, authors use Deep Belief Networks to pre-train a context-dependent artificial neuron net (ANN)/HMM system that was learned on two datasets. To enhance its voice recognition capabilities, Google created a front-end for neural networks.  For low-resource languages, the authors of \cite{besacier2014automatic} try to use Google Speech Recognition. Authors of \cite{lei2013accurate} from Google Inc. report the creation of an effective, compact, and extensive voice recognition system for mobile devices. HTK is used by Shelza Dua, Mohit Dua, R.K. Aggarwal, Virender Kadyan, among others \cite{dua2012punjabi} for Punjabi Automatic Speech Recognition.

It is advised to employ HTK and Julius in voice recognition instructional activities. Research tasks can be successfully conducted with Kaldi \cite{belenko2017comparative}. Examine deep neural networks (DNNs) with many hidden layers as well as the methods used to train them in \cite{hinton2012deep}. In \cite{jaitly2012application}, authors use Deep Belief Networks to pre-train a context-dependent artificial neuron net (ANN)/HMM system that was learned on two datasets. To enhance its voice recognition capabilities, Google created a front-end for neural networks.  For low-resource languages, the authors of \cite{besacier2014automatic} try to use Google Speech Recognition. Authors of \cite{lei2013accurate} from Google Inc. report the creation of an effective, compact, and extensive voice recognition system for mobile devices. HTK is used by Shelza Dua, Mohit Dua, R.K. Aggarwal, Virender Kadyan, among others \cite{dua2012punjabi} for Punjabi Automatic Speech Recognition.

For Arabic phonemes, \cite{el2016building} authors utilize CMU Sphinx. Convolutional neural networks are used for error rate reduction by the authors of \cite{abdel2014convolutional}. Techniques for creating noise-resistant voice recognition systems are analyzed by Jinyu Li, Li Deng, Yifan Gong, and Reinhold Haeb-Umbach \cite{li2014overview}. Oliver Lemon \cite{lemon2012data} and Jerome R. Bellegarda \cite{bellegarda2014spoken}, as well as Li Deng and Xiao Li \cite{deng2013machine}, investigate speech recognition interfaces, and from these articles, we may distinguish Siri. Silnov Dmitry Sergeevich \cite{silnov2016special} uses Yandex Speech-Kit and Google Speech to decode radio conversations.

The authors of \cite{smirnov2016russian} present their CMU Sphinx-based recognition system. The authors of \cite{kamarudin2013low} suggest using Microsoft Speech API to automate your home. The usage of Microsoft Speech SDK to provide an online resource for students to train their speaking abilities is covered by Howard Hao-Jan Chen \cite{chen2011developing}. Kaldi Speech Recognition Toolkit is described by Povey, D. el. \cite{lakkhanawannakun2019speech}. Microsoft Speech Engine is used in Ivan Tashev's \cite{tashev2013kinect} and R. Maskeliunas, K. Ratkevicius, and V. Rudzionis' \cite{maskeliunas2011voice} analysis of human-machine interaction. Microsoft Speech API is a suggestion made by Y. Bala Krishna, S. Nagendram \cite{krishna2012zigbee}, Faisal Baig, Saira Beg, and Muhammad Fahad Khan \cite{baig2013zigbee} for usage with smart home equipment. The authors of \cite{sharma2012speech} suggest using Microsoft Speech API to create assistive technology that will enable communication between two physically impaired people—the blind and the deaf. The writers of \cite{yamamoto2017presenter} employ the Microsoft Speech API to assess audience response.

The Microsoft Speech API is utilized by the authors \cite{rai2017efficient} to create an examination system that is accessible to students with disabilities. The use of Microsoft Speech API for discussion systems is also examined by the authors in \cite{kumar2017knowledge}. A language model on Holy Quran recitations was trained and evaluated using CMU Sphinx tools \cite{el2016towards}. The CMU Sphinx is used by the creators of \cite{phull2016investigation} to train the system and decipher voice data. Using Sphinx technologies, Hassan Satori and Fatima ElHaoussi \cite{satori2014investigation} create their own continuous automatic speech recognition system that is speaker-independent. 

The same authors demonstrate their ability to define Smokers and Nonsmokers utilizing their CMU Sphinx-based system. Sphinx is used by authors to create subtitles in \cite{kulkarni2016comprehensive} through a three-stepstep procedure that combines audio extraction, speech recognition, and subtitle synchronization. In \cite{plonkowski2014tuning} CMU Polish speech was recognized using Sphinx. I. Medennikov and A. Prudnikov \cite{medennikov2016advances} Speech by Kaldi Russian voice recognition tests were conducted using the recognition tool-set. The \cite{besacier2015speech} authors employ Kaldi for speech recognition in Africa. Kaldi is used by the authors in \cite{peddinti2016far} for speech recognition.

The Kaldi voice recognition system is being developed by authors \cite{yamamoto2017presenter} to be compatible with Julius. For the creation of a robot voice recognition system in \cite{sakai2015online}, the authors employ Julius. Julius is used by creators \cite{lojka2014multi} of mobile applications. HTK is used by the authors of \cite{matarneh2017speech} for recognition whisper. HTK was utilized in \cite{mankala2014automatic} for Telugu language recognition. HTK was also utilized in \cite{adetunmbi2016development} for their own speech-to-text system.

\subsection{Novelty}

Most of the current speech recognition models have several drawbacks for people with hearing, speech, or cognitive impairment. Speech recognition models may have trouble accurately identifying the speech of some people with speech impairments because they have trouble speaking. Using speech recognition models that have not been trained on a variety of accents and speech difficulties may also be challenging for people with accents or speech problems. Speech recognition algorithms might not be able to distinguish between non-standard speech, such as sign language or the alternate forms of speech used by those with hearing or speech impairments. It can be difficult for persons with hearing impairments to effectively understand speech in loud settings where speech recognition algorithms may struggle. Speech recognition algorithms may be problematic for people with cognitive disabilities since their speech may be slower, less distinct, or have different intonation patterns than those of average speakers.

The proposed model focused on working on the drawback of the present models and tried to enhance speech recognition models so that persons with hearing, speech, or cognitive impairments can use them more easily. The objective is to develop speech recognition models that are more user-friendly and capable of accurately recognizing a greater variety of speech patterns. Our model captures the speech from the user by the microphone, which then transmits the audio to the proposed model. First, the audio is converted into text. This conversation is done passing by the audio through two separate layers. Then the text which is produced is now transmitted to the morse code conversion layer, where the text is converted to its corresponding morse code. Now, this morse code can be conveyed to people with disabilities by using vibration or by various methods of touch. 

The proposed model of speech recognition have several advantage People with hearing, speech, or cognitive disabilities can communicate easily and effectively with Morse code, which can be especially helpful in emergency situations where other kinds of communication are inaccessible. Morse code is a suitable substitute for those who might find it difficult to use more complex forms of communication since it is a relatively simple system that is simple to learn and use.

\section{Proposed Work}
\label{sec:model}

\begin{figure}[h!]
  \includegraphics[width=\linewidth]{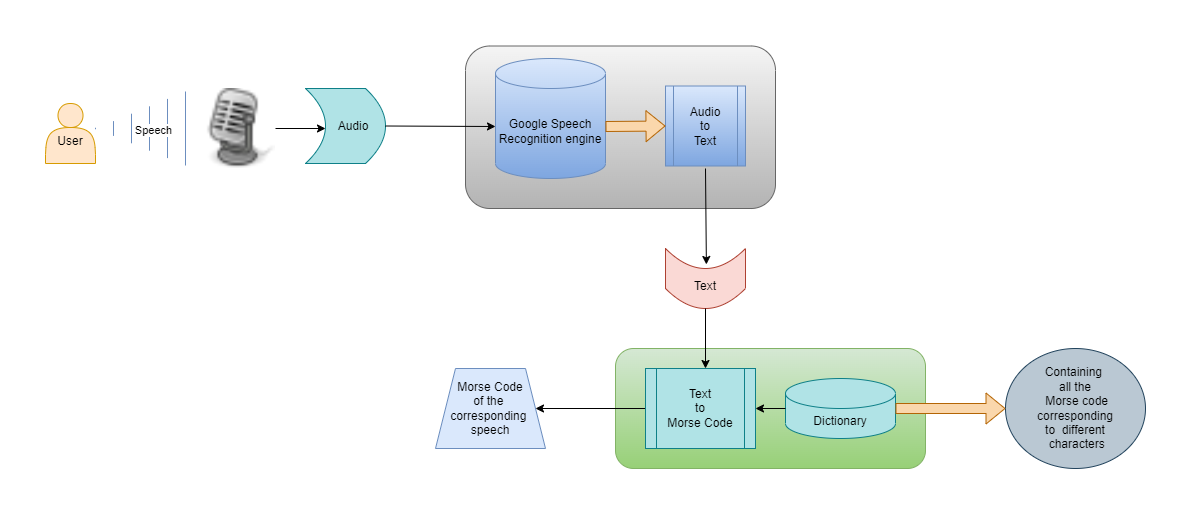}
  \caption{Working of our proposed model}
  \label{fig:working}
\end{figure}

Our speech recognition model focuses on making it useful for people with hearing, speech, or cognitive impairments. The proposed model collects speech from the user via a microphone. Microphones collect the speech and transmit the audio to the proposed model. After the proposed model receives the audio, it first converts it into text format. Then that text format is converted to morse code. The processing of the audio to the morse code and the workflow of the proposed model are shown in Figure \ref{fig:working}.

The audio file entering the proposed model is first passed through the speech recognition model (Section \ref{speech}) which converts the speech to text format. Within the speech recognition model (Section \ref{speech}); the audio is passed through a different neural network (section \ref{Acoustic} and \ref{Lan}) which in steps converts the speech to text. Once the speech is converted to text it is then transmitted to the morse code converter (Section \ref{morse}) which with the help of morse code dictionary (Section \ref{dictionary}) converts the text received to morse code

\subsection{\textbf{Speech Recognition : Speech to Text}}\label{speech}

\begin{figure}[h!]
  \includegraphics[width=\linewidth]{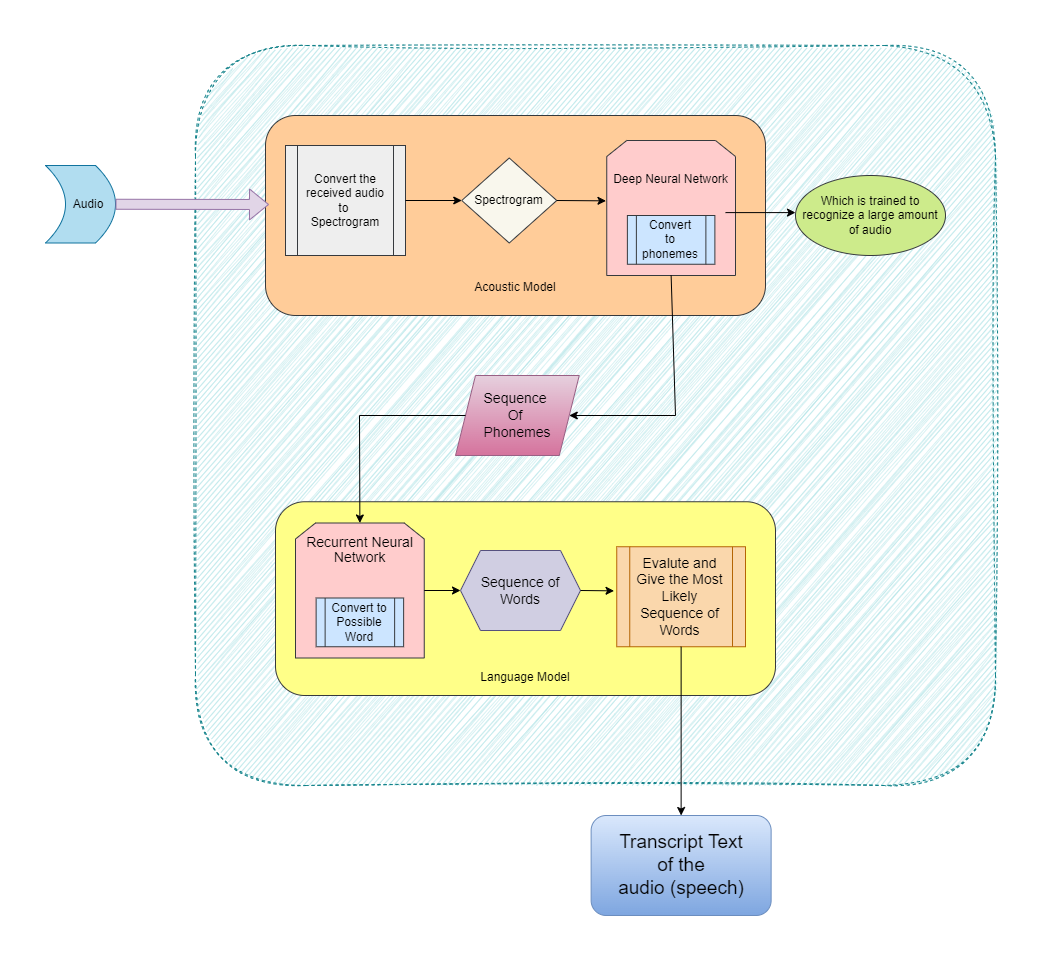}
  \caption{Working of the Speech Recognition}
  \label{fig:speech_model}
\end{figure}

The Speech Recognition technology utilized in the proposed model is a speech-to-text technology created by Google. In more than 80 languages and variants, it can convert spoken words into written text. Our proposed model's Speech Recognition accurately records speech using a combination of machine learning techniques and a lot of data. The technology transcribes speech using both acoustic and language models. A detailed working of this speech recognition technology is shown in Figure \ref{fig:speech_model}

The audio of the input speech is processed by the acoustic model of the speech recognizer of the proposed model and turned into a spectrogram. Spectrogram is a graphic representation of the audio. A neural network that has been trained to distinguish various sounds and phonemes is then provided with the spectrogram. This network produces a series of phonemes, which are the basic components of sound in a language. A detailed description about this acoustic model is done in \ref{Acoustic}.

The language model of the speech recognizer, on the other hand, interprets the phoneme sequence received from the acoustic model and turns it into a word sequence. The most likely word or phrase that the speaker pronounced is predicted by taking into consideration the context and grammatical rules of the language. A neural network that was trained using a sizable text dataset is also utilized by the language model.A detailed description of this Language Model is provided in
\ref{Lan}.

The system integrates the output of both models to create the final transcription after the acoustic and linguistic models have processed the audio. The system also employs a technique called beam search algorithm, which considers several word and phoneme combinations to choose the most probable transcription.

\subsubsection{Acoustic Model} \label{Acoustic}

The deep neural network (DNN) utilized in the speech recognition technology of our model was trained on a lot of audio data. The model's precise architecture and training information are confidential and not made available to the public by Google. However, in general, the model is most likely to be a multi-layered Time-Delay Neural Network (TDNN) or a variation of a Long Short-Term Memory (LSTM) network. A spectrogram or Mel-Frequency Cepstral Coefficients (MFCCs) of the audio are commonly used as the network's input, and its output is a probability distribution over a collection of potential phonemes or sub-words. To improve the network's parameters for precise voice recognition, the model is trained using a combination of supervised and unsupervised learning methods, including backpropagation and Connectionist Temporal Classification (CTC).

\subsubsection{Language Model} \label{Lan}

The speech recognition technology used in the proposed model uses a recurrent neural network (RNN) variation that has been trained on a lot of text data. However, in general, the model is to be a multi-layer transformer network or a variation of an Long Short-Term Memory (LSTM) network. The network typically receives a series of phonemes or sub-words produced by the acoustic model mention in \ref{Acoustic} as input, and the result is a probability distribution over the set of potential words or phrases. To optimize the network's parameters for precise speech recognition, the model is trained using supervised learning methods like backpropagation. The acoustic model's sequence of sub-words is evaluated using the trained model to assign probabilities, and the word sequence with the highest probability is selected as the transcription.

\subsection{\textbf{Morse Code Converter : Text to Morse Code}}\label{morse}

The suggested model's morse code converter is a function that accepts text input as input and outputs the appropriate morse code. The morse code dictionary and the function are both functional. The dictionary links every character and number in the English alphabet to its appropriate representation in Morse code. Initially, the converter initialises a blank string to hold the generated Morse code. 

Every letter in the given text is tokenized by the converter. It determines whether a given letter is a space or not for each one. In the event that it isn't a space, it adds the appropriate Morse code representation to the string of Morse code after locating it in the dictionary. It only adds a space to the morse text string if the letter is a space. 

After tokenizing through all the letters in the input text, the function returns the morse code string, which contains the complete Morse code representation of the input text. 

Let's consider an example, if the speaker says "HELLO" then the transmitted text to the converter is "HELLO", the converter will first tokenize each letter and look up the corresponding Morse code representation from the morse code dictionary and add it to the morse code string. So, the final output will be ".... . .-.. .-.. --- "

\begin{figure}[h!]
  \includegraphics[width=\linewidth]{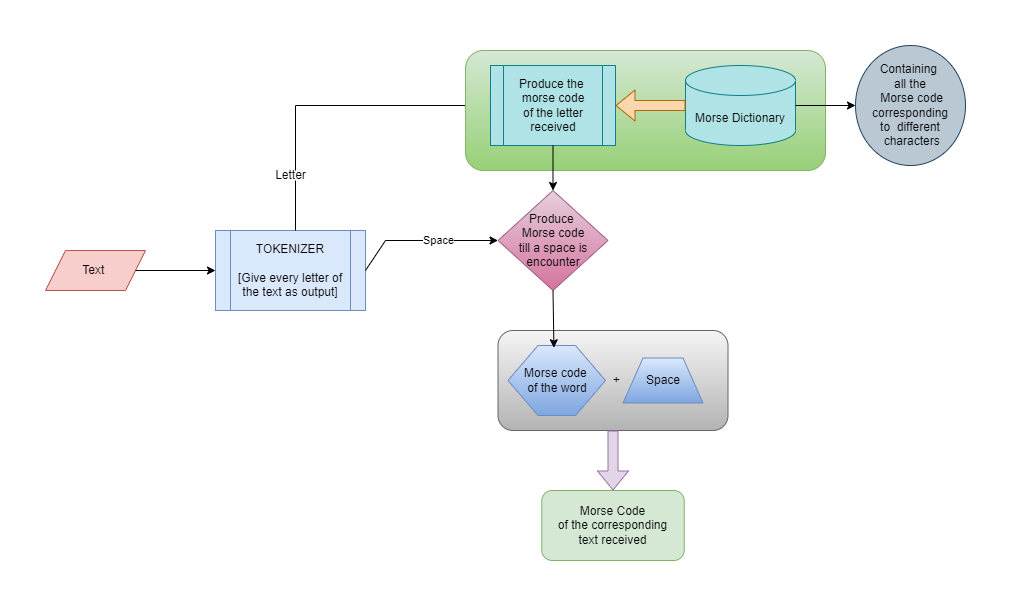}
  \caption{Working of the Morse Code Converter}
  \label{fig:morse}
\end{figure}

\subsubsection{Morse Code Dictionary} \label{dictionary}

The morse code dictionary is used in the morse code converter to convert each character of the transmitted text to its corresponding morse code representation. The key in the dictionary is the letters and numbers that are present in the transmitted text, and the values are the corresponding Morse code representations of those letters and numbers. By using this dictionary, the morse code converter is able to convert each character in the transmitted text to its corresponding Morse code representation, store the final result and give it as output.

\section{Results and Analysis}
\label{result}

The accuracy of the proposed model depends only on the accuracy of the speech recognition layer of our model. This dependency is because the morse code converter is a process and thus its accuracy is always 100\%. The morse code conversion layer's accuracy depends completely on the output generated by the speech recognition layer. 

The unit used in this study to measure the accuracy of the voice recognition layer is the word error rate (WER) \cite{gruhn2011statistical,chelba2012large}. The ratio of transcription errors to total words spoken is known as WER. A lower WER in speech-to-text indicates better voice recognition accuracy. The Levenshtein distance, frequently referred to as the edit distance, is the source of WER. It determines the minimum number of edit operations required to change one string to   obtain another \cite{stenman2015automatic}.

\begin{table}[h]
\centering
\caption{Accuracy of the speech recognition used in the proposed model}
\label{tab:accuracy}
\begin{tabular}{lllll}
\hline
Recorded Files& \begin{tabular}[c]{@{}l@{}}Total\\ Word\end{tabular} & \begin{tabular}[c]{@{}l@{}}Total\\ Faults\end{tabular} & WER     & \begin{tabular}[c]{@{}l@{}}Time\\ (ms)\end{tabular} \\ \hline

\begin{tabular}[c]{@{}l@{}}Short Sentence\\ (six or less words\\  per sentence)\end{tabular} & 462 & 26 & 5.62\%  & 2889.45 \\ \hline

\begin{tabular}[c]{@{}l@{}}Long Sentence \\ (seven or more words \\  per sentence)\end{tabular} & 1024 & 118 & 11.52\% & 4215.75\\ \hline

\begin{tabular}[c]{@{}l@{}}Short Sentence \\ with artificially\\ added noise\end{tabular}& 462 & 42 & 9.09\%  & 2962.44 \\ \hline

\begin{tabular}[c]{@{}l@{}}Long Sentence\\ with artificially\\ added noise\end{tabular}& 1024 & 130 & 12.69\% & 4495.19 \\ \hline
\end{tabular}
\end{table}

\begin{figure}[h!]
     \begin{minipage}{\linewidth}
         % \centering
         \includegraphics[width=\linewidth]{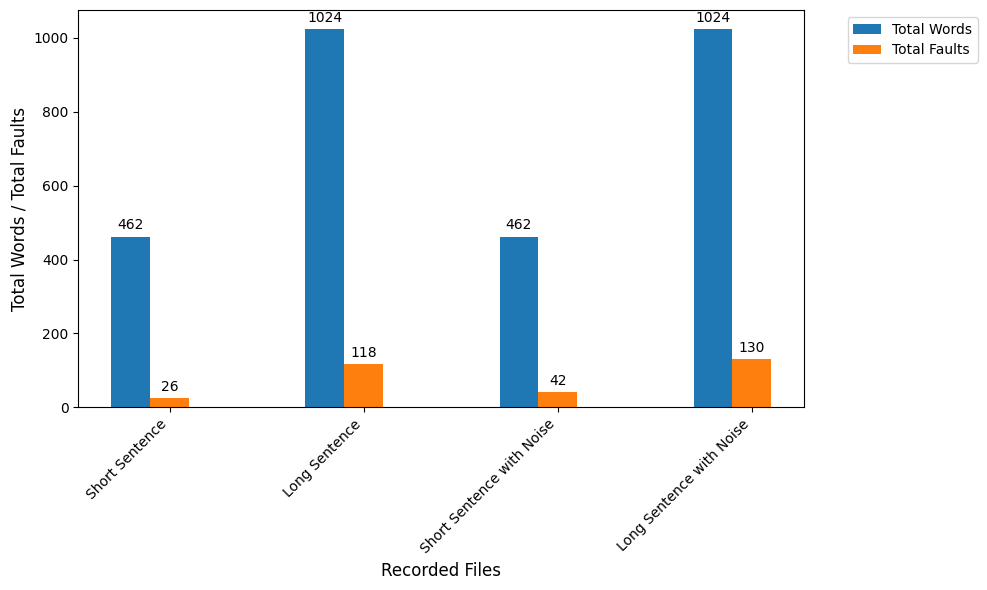}
          \caption{Accuracy of the speech recognition used in the proposed model}
        \label{fig:acc}
     \end{minipage}
     \hfill
     \begin{minipage}{\linewidth}
         % \centering
         \includegraphics[width=\linewidth]{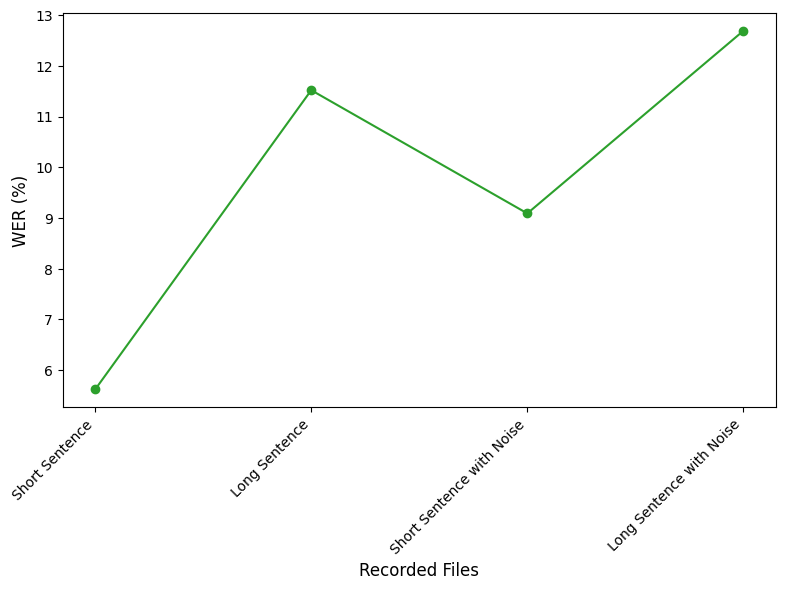}
         \caption{Variation of WER on different Recorded files}
  \label{fig:graph1}
     \end{minipage}
\end{figure}

\begin{figure}[h!]
     \begin{minipage}{\linewidth}
         % \centering
         \includegraphics[width=\linewidth]{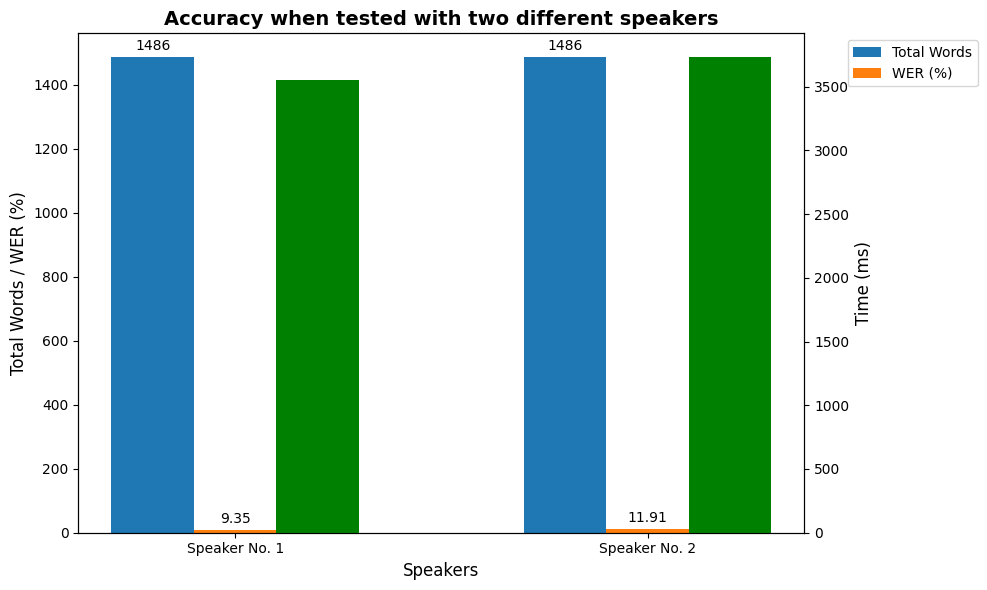}
         \caption{Accuracy when tested with two different speakers}
  \label{fig:graph2}
     \end{minipage}
\end{figure}

Table \ref{tab:accuracy} shows the WER percentage of the speech recognition used in the proposed model and the mean translation time in milliseconds when tested with recorded files with different parameters represented in figure \ref{fig:acc}. Figure \ref{fig:graph1} shows the graphical variation of the WER of the proposed model when tested with different recorded files. Table \ref{tab:speaker} shows the comparisons between WER when tried with two different system speakers tested with the same recorded files listed in table \ref{tab:accuracy}.Figure \ref{fig:graph2} shows the graphical variation of the WER of the proposed model when tested on the same different recorded files but with different system speakers. From the tables \ref{tab:accuracy} and \ref{tab:speaker}, we can conclude that the average WER of the speech recognition is 10.18\% and therefore the accuracy of the speech recognition layer of the proposed model is 89.82\%.

\begin{table}[h!]
\centering
\caption{Accuracy when tested with two different speaker}
\label{tab:speaker}
\begin{tabular}{llllll}
\hline
\begin{tabular}[c]{@{}l@{}}System\\ Speaker\end{tabular} & \begin{tabular}[c]{@{}l@{}}Total \\ Word\end{tabular} & \begin{tabular}[c]{@{}l@{}}Total\\ Fault\end{tabular} & WER     & Accuracy & \begin{tabular}[c]{@{}l@{}}Time\\ (ms)\end{tabular} \\ \hline
\begin{tabular}[c]{@{}l@{}}Speaker \\ No. 1\end{tabular} & 1486                                                  & 139                                                   & 9.35\%  & 90.65\%  & 3549.575                                            \\ \hline
\begin{tabular}[c]{@{}l@{}}Speaker\\ No. 2\end{tabular}  & 1486                                                  & 177                                                   & 11.91\% & 88.09\%  & 3731.840                                            \\ \hline
\end{tabular}
\end{table}

However, the assessment parameters and therefore the proposed model's accuracy can change in a variety of ways \cite{tebelskis1995speech} : 

\begin{enumerate}
  \item Confusable words and vocabulary size : A smaller vocabulary makes it simpler for the system to identify the right term than a bigger one. With a larger vocabulary, error rates inevitably rise. For instance, it is possible to recognize the numerals 0 through 10 properly, but mistake rates rise with larger vocabulary sets or the inclusion of confusable terms, or words with similar sounds. For instance, the phrases dew and you sound extremely similar yet have very different meanings.
  
  \item Speaker independence vs. dependence: Depending on the speaker and training, a speaker-dependent system is usually more accurate than a speaker-independent system. Furthermore, there are speaker-adaptive systems and multi-speaker systems that are intended for small user groups and that need very little speech data for training and understanding any speaker.
  
  \item Isolated, discontinuous, or continuous speech : The easiest to identify are isolated, which refers to single words, and discontinuous, which refers to complete sentences with artificially separated words by silence. Because of co-articulation and hazy borders, continuous speech is the most challenging to identify, but it's also the most fascinating because it lets us speak normally.
  
  \item Task and language constraints : The restrictions may be job-specific, admitting only statements that are pertinent to the activity at hand, such as "The car is blue" being rejected by a business that sells tickets. Others may be syntactic, rejecting "Car sad the is," or semantic, rejecting "The car is sad." Grammar represents constraints by removing absurd statements, and the perplexity of a sentence—a measure of the grammar's branching factor, or the number of possible words after a given word—measures how perplexing a sentence is.
  
  \item Read speech vs. Spontaneous : Compared to spontaneous speech, which can include words like "uh" and "um," as well as stuttering, coughing, and laughter, read speech from a text is simple to understand.
  
  \item Recording conditions : Background noise, acoustics (such as echoes), microphone type (such as close-speaking, telephone, or omnidirectional), frequency bandwidth limitations (such as telephone communications), and changing speaking styles all have an impact on performance (shouting, speaking quickly, etc.).
  
\end{enumerate}

To compare the speech recognition performance of our proposed model with other state-of-the-art speech recognition models/APIs, we conducted an experiment similar to that presented in \cite{herchonvicz2019comparison}. Specifically, we evaluated the performance of our model against the Bing Speech API and IBM Watson Speech to Text (STT) using a dataset comprising 20 paragraphs, each consisting of 30 sentences. In total, the models were tested on 300 sentences, and their performances were compared based on accuracy and the number of correctly recognized sentences.

\begin{figure}[h!]
     \begin{minipage}{\linewidth}
         % \centering
         \includegraphics[width=\linewidth]{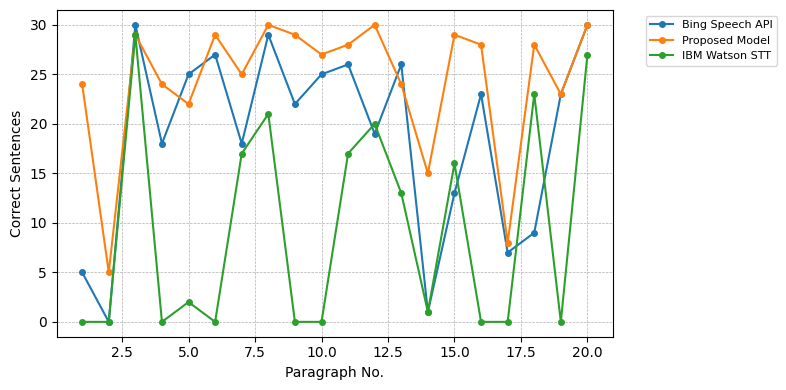}
          \caption{Variation by Different Speech Recognition Engines}
        \label{fig:collect}
     \end{minipage}
\end{figure}

\begin{table}[h!]
\centering
\caption{Comparison of Correct Sentences by Different Speech Recognition Engines}
\label{tab:comparison}
\begin{tabular}{llll}
\hline
\textbf{\begin{tabular}[c]{@{}l@{}}Paragraph\\ Number\end{tabular}} & \textbf{\begin{tabular}[c]{@{}l@{}}Bing Speech\\ API\end{tabular}} & \textbf{\begin{tabular}[c]{@{}l@{}}Proposed\\ Model\end{tabular}} & \textbf{\begin{tabular}[c]{@{}l@{}}IBM Watson\\ STT\end{tabular}} \\
\hline
1  & 5  & 24 & 0  \\ \hline
2  & 0  & 5  & 0  \\ \hline
3  & 30 & 29 & 29 \\ \hline
4  & 18 & 24 & 0  \\ \hline
5  & 25 & 22 & 2  \\ \hline
6  & 27 & 29 & 0  \\ \hline
7  & 18 & 25 & 17 \\ \hline
8  & 29 & 30 & 21 \\ \hline
9  & 22 & 29 & 0  \\ \hline
10 & 25 & 27 & 0  \\ \hline
11 & 26 & 28 & 17 \\ \hline
12 & 19 & 30 & 20 \\ \hline
13 & 26 & 24 & 13 \\ \hline
14 & 1  & 15 & 1  \\ \hline
15 & 13 & 29 & 16 \\ \hline
16 & 23 & 28 & 0  \\ \hline
17 & 7  & 8  & 0  \\ \hline
18 & 9  & 28 & 23 \\ \hline
19 & 23 & 23 & 0  \\ \hline
20 & 30 & 30 & 27 \\ \hline
\textbf{Total} & \textbf{376/600} & \textbf{487/600} & \textbf{186/600} \\
\hline
\end{tabular}
\end{table}

The results of our evaluation indicate that the proposed model achieved the highest performance, with an average accuracy of 0.978. The Bing Speech API followed closely with an average accuracy of 0.945, while IBM Watson STT had an average accuracy of 0.881. These results underscore the superior accuracy of our proposed model in recognizing speech as shown in figure \ref{fig:collect}.

\begin{figure}[h!]
     \begin{minipage}{\linewidth}
         % \centering
         \includegraphics[width=\linewidth]{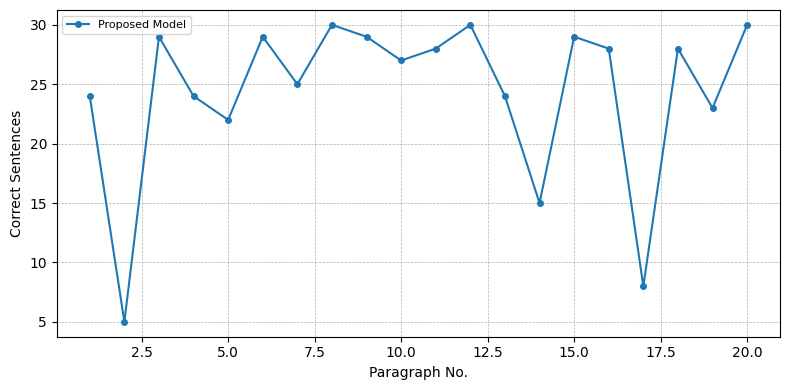}
          \caption{Proposed Model Correct Sentences}
        \label{fig:sep}
     \end{minipage}
     \hfill
     \begin{minipage}{\linewidth}
         % \centering
         \includegraphics[width=\linewidth]{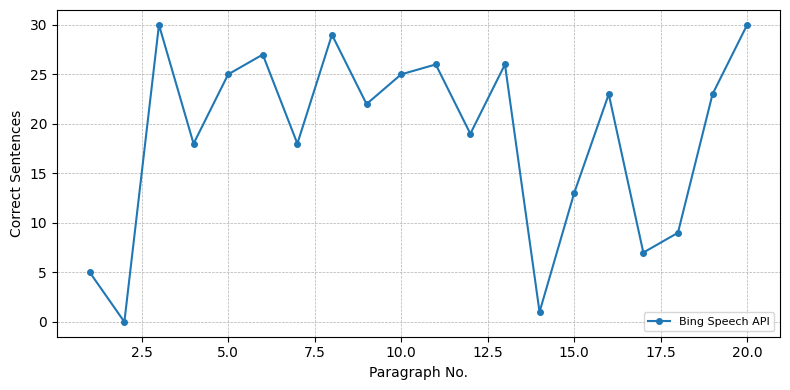}
          \caption{Bing Speech API Correct Sentences}
        \label{fig:sep1}
     \end{minipage}
     \hfill
     \begin{minipage}{\linewidth}
         % \centering
         \includegraphics[width=\linewidth]{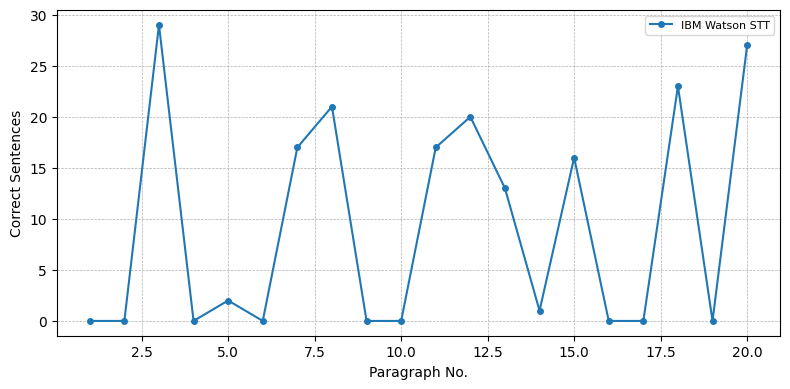}
         \caption{IBM Watson STT Correct Sentences}
  \label{fig:sep2}
     \end{minipage}
\end{figure}

Table \ref{tab:comparison} presents the evaluation of the correct answers provided by each speech recognition engine. The proposed model demonstrated a clear advantage, correctly recognizing 487 out of 600 sentences, which corresponds to an accuracy of approximately 81\% represented in Figure \ref{fig:sep}. The Bing Speech API achieved 376 correct sentences out of 600, representing an accuracy of about 62\% represented in Figure \ref{fig:sep1}. IBM Watson STT ranked last, with 186 correct sentences out of 600, translating to an accuracy of roughly 31\% represented in Figure \ref{fig:sep2}.

\begin{figure}[h!]
     \begin{minipage}{\linewidth}
         % \centering
         \includegraphics[width=\linewidth]{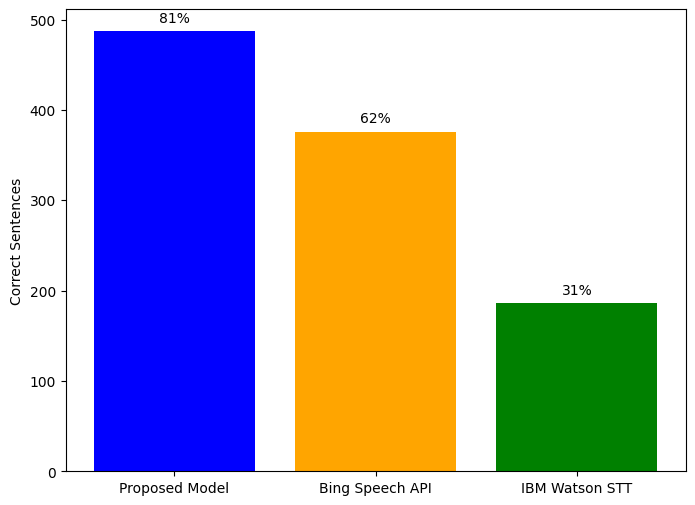}
          \caption{Evaluation of Correct Sentences by Each Speech Recognition Engine}
        \label{fig:bar}
     \end{minipage}
\end{figure}

These findings as summarized in table \ref{tab:sentence}  and figure \ref{fig:bar} highlight the effectiveness of the proposed model in terms of both accuracy and reliability, suggesting its potential for various applications requiring robust speech recognition capabilities. Future work will involve further validation of the model across more diverse datasets and in real-world scenarios to ensure its generalizability and practical utility. Additionally, improvements in the algorithm and architecture of the proposed model will be explored to enhance its performance further.

The comparative analysis illustrates that our proposed model not only surpasses the Bing Speech API and IBM Watson STT in terms of accuracy but also demonstrates a higher capability in recognizing a greater number of sentences correctly. This study contributes to the ongoing advancements in speech recognition technology by providing empirical evidence of the proposed model's superior performance. Future research will aim to refine the model further and explore its applicability in a broader range of use cases.

\begin{table}[h!]
\centering
\caption{Evaluation of Correct Sentences by Each Speech Recognition Engine}
\label{tab:sentence}
\begin{tabular}{l l l}
\hline
\textbf{Speech Recognition Engine} & \textbf{Correct Sentences} & \textbf{Percentage (\%)} \\
\hline
Proposed Model            & 487/600           & 81\%            \\
\hline
Bing Speech API           & 376/600           & 62\%            \\
\hline
IBM Watson STT            & 186/600           & 31\%            \\
\hline
\end{tabular}
\end{table}

\section{Benefit to the Society}
\label{benefit}

The proposed model takes audio speech from the user's audience as input. This audio speech can be in the form of spoken words, phrases, sentences, or any vocal communication. The model processes and analyzes this input for various purposes, such as transcription, sentiment analysis, speech recognition, or any other relevant task. The user's audience refers to the individuals or group of people who are speaking or delivering the audio content. 

The model is designed to work with this audio data and derive meaningful information or insights from it, depending on the specific application or use case. In essence, the model serves as a tool for processing and making sense of spoken language in audio form, providing valuable functionality and analysis based on the input it receives from the user's audience. 

The proposed model is also designed to provide a unique and innovative way of conveying information to the user. Specifically, it takes spoken speech as its input and converts it into Morse code, a series of dots and dashes representing letters and numbers in a symbolic fashion. However, instead of displaying this Morse code output through traditional means such as sound or visual cues, it utilizes a sensory interface located on the lower surface of the user's arm. 

In this context, the lower surface of the user's arm serves as a tactile communication platform. As the model processes the spoken speech, it translates it into the corresponding Morse code sequences, which are then conveyed to the user through a series of tactile sensations or vibrations on their arm. Each dot and dash in Morse code is represented by a distinct tactile signal, allowing the user to feel and interpret the converted message through their sense of touch.

 The proposed model offers a myriad of significant advantages for individuals facing challenges related to hearing, speech, or cognitive disabilities. Its core strength lies in providing an accessible, versatile, and universally comprehensible mode of communication, which can result in heightened independence, bolstered self-confidence, and improved self-esteem. 
 
The Morse code model transcends barriers and offers a transformative approach to communication for individuals with hearing, speech, or cognitive disabilities, with the potential to significantly enhance their quality of life and participation in society.

 In summary, the Morse code model holds profound societal advantages, particularly in its capacity to enhance inclusivity and accessibility for individuals grappling with hearing, speech, or cognitive disabilities. It serves as a transformative tool that has the potential to bridge longstanding gaps between those with disabilities and the wider community.

\section{Conclusion}
\label{conclusion}

We created a speech recognition system and reviewed it. The primary contribution of our work is that the proposed model focus on making speech recognition more accessible to people with hearing, speech, or cognitive impairments. Our method tries to develop a technology completely useful for only hearing, speech, or cognitive disabled people.

The proposal is mainly divided into two parts: Speech recognition and morse code conversion. The speech recognition receives the speech from the user. Then it converts the audio to text. This text produced acts as the input of the morse code conversion layer. It is where the text is converted to morse code and it is the final output of the proposed model.

Other than helping people with hearing, speech, or cognitive disabilities, our model can also be used for military purposes. Morse code can employed in the military for telegraph transmission as well as for communication between ships and ground forces and headquarters. It was widely taught to military troops and became a necessary instrument for communication during military operations because to its simplicity and dependability. Morse code can also be beneficial when verbal communication is impossible because it can be sent through a variety of techniques (e.g., flashing lights, horns). 

\section{Future Scope}

By raising the speech recognition layer's accuracy, the suggested model may be enhanced. Our model can be rebuilt and developed further in the future and put it in work in a better manner. Now; if the morse code conversion layer is replaced with the Braille script conversion layer which converts the text produced from the speech recognition layer to braille script. Thereby it will come handy for the people with hearing and visual disabilities and also it will come as a helpful tool for people who are already trained or going to get trained in braille script. 

Further more, as mentioned earlier in Section \ref{conclusion}, morse code can play a very helpful role in various military purposes. Therefore the proposed model can be rebuild and redesigned for any such specific purposes. There might be other scopes where morse can be put to use, and therefore the model can be redesign for that purpose.

\bibliographystyle{IEEEtran}
\bibliography{reference}

\end{document}